\documentclass[10pt, conference]{IEEEtran}
\usepackage[T1]{fontenc}% optional T1 font encoding
\usepackage{color}
\usepackage{soul}
\usepackage{caption}
\usepackage{multirow}
\usepackage{cite}
\usepackage{pdfpages}
\usepackage{epstopdf}
\usepackage[cmex10]{amsmath} % Math Package
\usepackage{amssymb}
\usepackage{algorithm}
\usepackage{algorithmic}

\usepackage{array}
\usepackage{amsmath}
\usepackage[tight,footnotesize]{subfigure}
\usepackage{url} % PDF, URL, and Hyper Link Package
\usepackage{multirow}
\usepackage[inline]{enumitem}
\usepackage[dvipsnames]{xcolor}
\usepackage{makecell}
\usepackage{url}

\usepackage{amsmath}
\usepackage[cmintegrals]{newtxmath}
\usepackage{blindtext}

\begin{document}
\title{Traffic-Aware Dynamic Functional Split for 5G Cloud Radio Access Networks}

% \author {Himank Gupta,
%          Mayank Singh, 
%          Antony Franklin A, 
%          and  Bheemarjuna Reddy Tamma 
%          %<-this % stops a space
% }
\author{\IEEEauthorblockN{Himank Gupta\IEEEauthorrefmark{1}, Antony Franklin A\IEEEauthorrefmark{1}, 
Mayank Kumar\IEEEauthorrefmark{2} and
Bheemarjuna Reddy Tamma\IEEEauthorrefmark{1}}
\IEEEauthorblockA{Indian Institute of Technology Hyderabad, India\IEEEauthorrefmark{1},
Indian Institute of Technology Bhilai, India\IEEEauthorrefmark{2}\\
Email: \IEEEauthorrefmark{1}cs16mtech01001@iith.ac.in
\IEEEauthorrefmark{1}antony.franklin@iith.ac.in,
\IEEEauthorrefmark{2}mayankk@iitbhilai.ac.in,
\IEEEauthorrefmark{1}tbr@iith.ac.in}}

\maketitle

\begin{abstract}
% \vspace{-0.2cm}
The recent adaption of virtualization technologies in the next generation mobile network enables 5G base station to be segregated into a Radio Unit (RU), a Distributed Unit (DU), and a Central Unit (CU) to support Cloud based Radio Access Networks (C-RAN). RU and DU are connected through a fronthaul link. In contrast, CU and DU are connected through a midhaul link. Although virtualization of CU gives benefits of centralization to the operators, there are other issues to be solved such as optimization of midhaul bandwidth and computing resources at edge cloud and central cloud where the DUs and CUs are deployed, respectively. In this paper, we propose a dynamic functional split selection for the DUs in 5G C-RAN by adopting to traffic heterogeneity where the midhaul bandwidth is limited. We propose an optimization problem that maximizes the centralization of the C-RAN system by operating more number of DUs on split Option-7 by changing the channel bandwidth of the DUs. The dynamic selection of split options among each CU-DU pair gives 90\% centralization over the static functional split for a given midhaul bandwidth. 
\end{abstract}

\begin{IEEEkeywords}
\vspace{-0.1cm}
5G, dynamic functional split, fronthaul, midhaul, virtualisation.
\vspace{-0.3cm}
\end{IEEEkeywords}

\IEEEpeerreviewmaketitle
\vspace{-0.18cm}
\section{Introduction }
\vspace{-0.18cm}
The recent exponential growth of network users and enormous capacity requirements of mobile applications such as high definition video streaming and AR/VR videos imposes high requirements on future networks. In anticipation of such high need, Mobile Network Operators (MNOs) and researchers have investigated various solutions such as Massive multiple-input multiple-out (MIMO), Millimeter-wave (mmWave) communications, or deploying small cells to offload the load of macrocells. Each of these solutions has its limitations. Some challenges related to channel estimation, user scheduling, energy efficiency, and deployment cost have been discussed by authors in~\cite{mimo}.
Therefore, Cloud Radio Access Network (C-RAN) has emerged as a prominent solution that can support up to hundreds of gigabit data rates cost-effectively without degrading the performance~\cite{costcran}. In recent times, industries are adopting C-RAN architecture for the 5G network~\cite{mavenir} due to less CAPital EXpenditure, and OPerational EXpenditure (CAPEX \& OPEX) in deployment and better scalability. C-RAN is considered as an architectural solution that can reduce the CAPEX \& OPEX in dense 5G cellular networks while allowing better network performance. 

% In anticipate of such high needs, cloud-radio access networks (C-RANs) has been designed that can support up to hundreds of gigabit data rates in a cost-efficient way without degrading the performance. C-RAN is the network architecture for cellular systems, in which conventional base station is separated into the baseband processing unit (BBU) and the RF antenna unit called Remote Radio Head (RRH) in a manner that the BBUs are centralized in a BBU pool and the RRHs are distributed in each service cell. The BBU is a baseband processing device of a conventional base station, and the RRH is an RF antenna unit of a conventional base station. A fronthaul network is formed between the BBU and the RRHs. Since the BBUs are centralized in a cloud data center, it is easy to operate, maintain and ensure cooperative communication in networks. C-RAN is considered as an architectural solution that can reduce the capital and operational expenditure in dense 5G cellular networks while allowing better network performance. 

\begin{figure}[t]
\centering
\includegraphics[width=0.9\columnwidth, keepaspectratio]{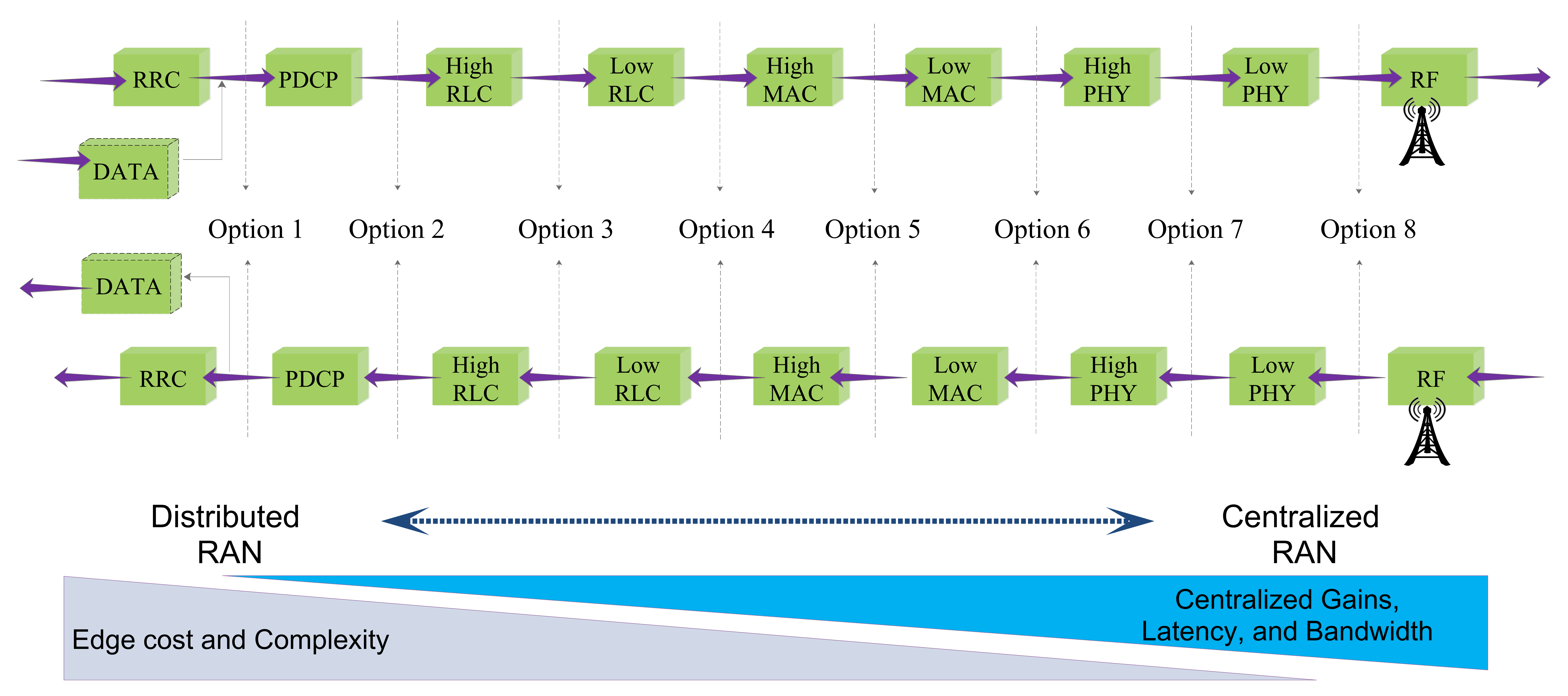}
 \caption {Split options as defined in the 3GPP TR 38.801.}
\label{figure:split}
\vspace{-0.74cm}
\end{figure}

In 5G C-RAN, the Base Station (BS) (also known as gNB) protocol stack is divided (\emph{i.e.,} it is functionally split) into the following network components~\cite{Himank}: 
\begin{enumerate*}[label=(\roman*)]
    \item the Central Unit (CU) which consists of the upper layers of the protocol stack such as Packet Data Convergence Protocol (PDCP) and Radio Resource Control (RRC);
    \item the Distributed Unit (DU) which consists of the lower layers of the protocol stack such as Radio Link Control (RLC),  Medium Access Control (MAC), and Upper Physical (U-PHY) layer;
    \item the Radio Unit (RU) which consists of the Lower Physical (L-PHY) layer functionalities. 
\end{enumerate*} 
The RU and DU communicate over a fronthaul interface (also called as fronthaul~I) while DU and CU communicate over a midhaul interface (also called as fronthaul~II). The fronthaul requirement depends on the functional split selected and network bandwidth used.
As shown in Fig.\ref{figure:split}, the 3GPP has proposed various functional splits~\cite{3gpp38801}, and the choice of optimal functional split by MNOs depends on several factors related to radio network deployment scenarios, traffic constraints, and intended supported services. Moreover, computationally costly operations like Fast Fourier Transformation (FFT), Inverse Fast Fourier Transformation (IFFT), Rate Matching, and Turbo encoding/decoding are shifted to the CU side as we move towards Option-8 from Option-1.
 MNOs intend to use Option-7 or Option-8 to gain maximum centralization and other benefits at the cost of more network bandwidth for the midhaul interface. Authors in~\cite{aptran} proposed a flexible functional split solution in which authors choose the different split options for each CU-DU pair based on the resource available. Due to limited resources, only a few CU-DU pairs can operate on Option-7 or Option-8. We accentuate that, the requirement of high network bandwidth for Option-7 or Option-8 primarily depends upon transferring IQ (Inphase and Quadrature) samples and other signaling information from DU to CU and vice-versa. Furthermore, the number of IQ  samples generated is based on channel bandwidth used by BS. Dynamically tuning the channel bandwidth is a viable option that supports the above proposition in the 5G C-RAN architecture. Based on available computing resources, most of the existing literature optimize the delay experienced in fronthaul or midhaul link~\cite{delay}. To the best of our knowledge, none of the recent works take into account the channel bandwidth in C-RAN. Hence, to mitigate challenges incurred due to high channel bandwidth in 5G C-RAN, we propose a Traffic-Aware Dynamic Functional Split that inculcates the channel bandwidth as one of the key parameters. The major contributions are summarized as follows: 
\begin{itemize}
    \item A dynamic functional split for 5G C-RAN is proposed to leverage midhaul bandwidth, and resulting maximum centralization by efficiently tuning channel bandwidth considering spatio-temporal traffic heterogeneity.
    \item An optimization  problem is proposed to optimally select functional split for a given traffic load at a given time. 
\end{itemize}

\vspace{-0.19cm}
\section{Traffic aware Dynamic Functional Split Selection}
\vspace{-0.15cm}
The bandwidth and latency budget required to run a fully centralized 5G C-RAN is very high, \emph{e.g.,} Option-8 of the 5G C-RAN with $100\ MHz$ channel bandwidth and $32$ antennae requires a midhaul bandwidth of $157.3\ Gbps$~\cite{3gpp_161813}. Such a high bandwidth is neither cost-efficient nor energy-efficient for MNOs. Moreover, deploying isolated gNBs is also not an adequate solution to meet the data rate requirement of the 5G network~\cite{Matching}. According to the 3GPP, 5G RAN can support a downlink peak data rate of up to $4\ Gbps$ using a channel bandwidth of $100\ MHz$~\cite{3gpp_162102}. Due to diurnal human activity pattern, the spatio-temporal traffic heterogeneity (tidal traffic) results in non-uniform utilization of available channel resources at the BS~\cite{kora}. The BS suffers from resource shortage during peak hours, and resources remain idle during non-peak hours. Hence, a dynamic scheme to efficiently utilize  channel bandwidth coping with varying traffic load is of utmost importance.

%The peak load of any BS occurs only for a few hours in a day based on the spatio-temporal traffic pattern. So, it is not required for the BS to use the entire available channel bandwidth all the time. So, when the BS traffic is below the peak downlink rate, we can use lower channel bandwidth according to the traffic load of the BS. 

% The following table shows the downlink data rate comparison between LTE and 5G. Different channel bandwidth at DU influences the different peak downlink rate.\\

\begin{table}[b]
\vspace{-0.29cm}
\footnotesize
    \begin{center}
\footnotesize
\begin{tabular}{|p{1.2cm}|c|c|c|}%{\columnwidth}
% \centering
\hline
\textbf{\makecell{Split\\ Option}} & \textbf{\makecell{Channel BW\\ $MHz$}} & \textbf{\makecell{Midhaul BW\\ $Gbps$}} &  \textbf{\makecell{Max DL Traffic\\ $Gbps$}} \\
\hline
\hline
\makecell{Option-2} & \makecell{$100$} &  \makecell{$UT + 0.016$}&  \makecell{$4$} \\
\hline
\makecell{Option-6} & \makecell{$100$} & \makecell{$UT + 0.133$}&  \makecell{$4$}\\
%  &  &  User traffic + 133Mb/s & 4Gbps \\
\hline
\multirow{5}{*}{\makecell{Option-7}} & $100$ & $9.4$ & $4$\\
                     & $80$ & $7.52$ & $3.2$\\
                     & $60$ & $5.64$ & $2.4$\\
                     & $40$ & $3.76$ & $1.6$\\
                     & $20$ & $1.88$ & $0.8$\\
                     \hline
\multirow{5}{*}{\makecell{Option-8}} & $100$ & $157.28$ & $4$ \\
                     & $80$ & $125.8$ & $3.2$\\
                     & $60$ & $94.37$ & $2.4$\\
                     & $40$ & $62.9$ & $1.6$\\
                     & $20$ & $31.45$ & $0.8$\\
\hline
\end{tabular}
\end{center}
    \caption{Maximum downlink traffic and midhaul bandwidth requirement for different channel bandwidths of 5G RAN}
    \label{midhaul}
    
\end{table}

\begin{figure}[t]
\centering
\includegraphics[width=0.9\columnwidth, keepaspectratio]{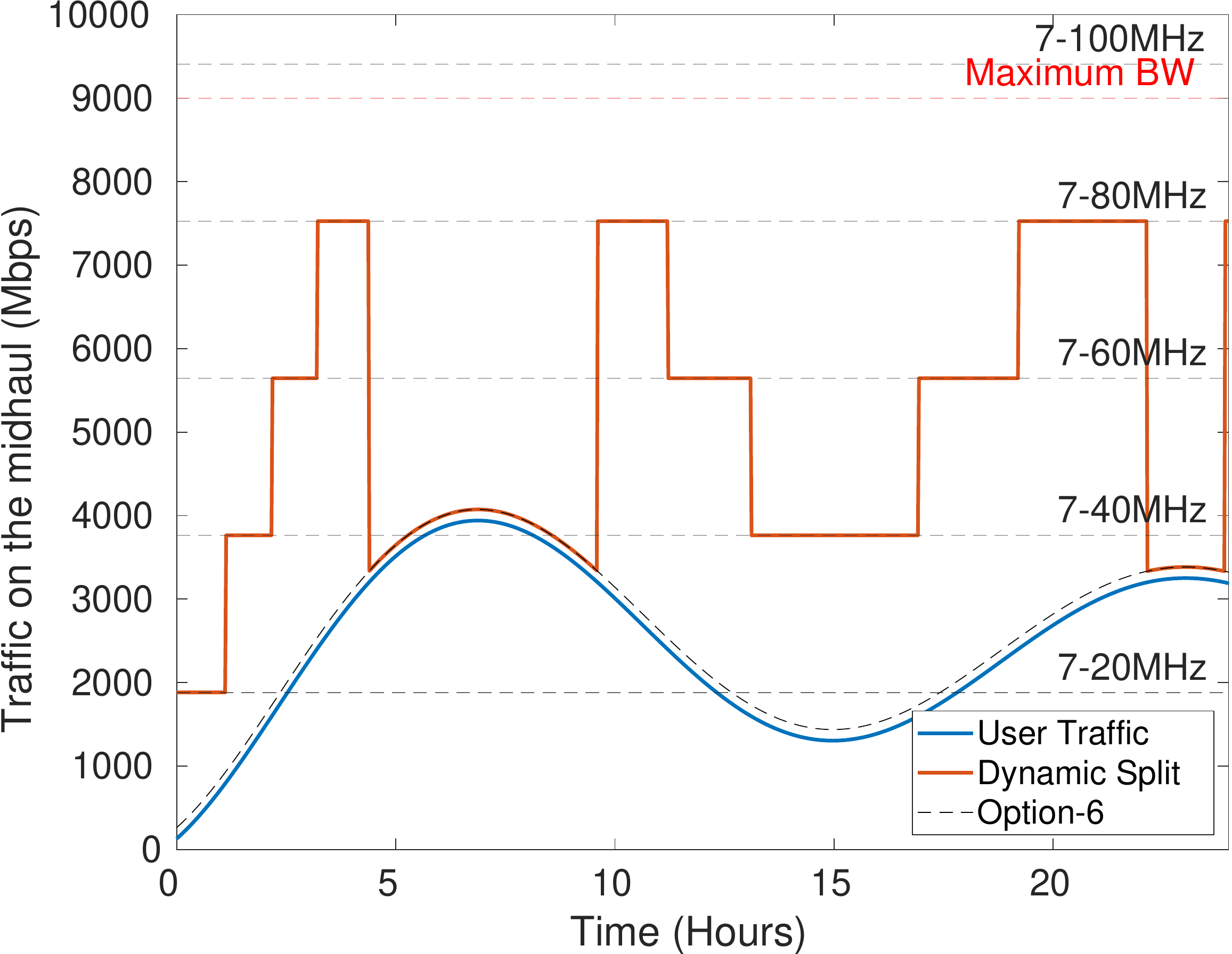}
 \caption {Midhaul traffic generated by 5G C-RAN implementing a traffic aware dynamic functional split.}
\vspace{-0.71cm}
\label{figure:analysis_II}
\end{figure}

Table~\ref{midhaul} shows the peak downlink traffic and midhaul bandwidth requirement for different functional split options for difference channel bandwidths~\cite{3gpp_162102}. From the table, we want to accentuate that Option-7 and Option-8 are traffic independent functional splits, \emph{i.e.,} midhaul bandwidth requirements for these options do not change based on the BS load. On the other hand, the bandwidth requirement for Option-2 and Option-6 changes with BS load \emph{i.e.,} the midhaul bandwidth requirement is the User Traffic ($UT$) from the BS with an added fixed overhead based on the split option used. These key features enable us to perform dynamic functional split in 5G C-RAN to achieve maximum centralization for a given midhaul bandwidth. Fig.~\ref{figure:analysis_II} shows an example scenario to demonstrate the benefits of using dynamic functional split based on channel bandwidth. The maximum available midhaul bandwidth is assumed to be $9\ Gbps$ in the figure. When the traffic changes, the dynamic split technique can switch from Option-7 $20\ MHz$ to Option-7 $80\ MHz$ to support the traffic load beyond which the midhaul bandwidth becomes the bottleneck. At this point, the BS can switch to Option-6 $100\ MHz$. Here, we do not consider Option-8, because for a minimum channel bandwidth (\emph{i.e.,} $20\ MHz$), it requires $31.45\ Gbps$ midhaul bandwidth for a one CU-DU pair. We are assuming that the operator does not have enough resources to provide such a high bandwidth to each CU-DU pair. Moreover, our model scrutinizes either Option-2 or Option-6 due to similar midhaul bandwidth requirement for these options. We can observe that, the dynamic functional split technique can give more centralization which in turn can benefit from Coordinated Multi-Point (CoMP), Inter-Cell Interference Coordination (ICIC)~\cite{load}, and energy efficiency by adapting to change in the BS load. In this paper, we propose an optimization model that can result best possible functional split for BSs of 5G C-RAN with an objective of maximizing the centralization for a given midhaul bandwidth. 
% Based on this, we can dynamically select the functional split for all the BSs in the network whenever the load on the BSs changes. 

\section{Optimal Functional Split as an Optimization Model}
\vspace{-0.15cm}
%\underline{Input}:
Let $\mathcal{B}$ be an array of different channel bandwidths in descending order which represents the breakpoints for Option-7 functional split (\emph{i.e.,} $100\ MHz$, $80\ MHz$, $60\ MHz$, etc.). The operator is assumed to use aforementioned set of bandwidths for the BSs. In our model, we assume that this set is same across all the BSs in the network. But the model can be extended for different set of bandwidths for each of the BSs as well. A glossary of mathematical notations used in optimization model are highlighted in Table~\ref{variables}. For each split option $i$, we can calculate the midhaul bandwidth cutoff $W(i)$ using Eqn.(\ref{eq:bandwidths}). Here, $i=1$ means Option-6, while $i>1$ means Option-7 with different channel bandwidth given in set $\mathcal{B}$.  

\begin{table}[h]
    \centering
    \footnotesize
   \begin{tabular}{|l|l|}
   \hline
        \textbf{Notation} & \textbf{\hspace{50pt} Definition}
        \\
    \hline   
    \hline
        $UT_i$ & User traffic at $DU_i$ in Mbps \\
         \hline
        $n_{DU}$ & Number of DUs\\
        \hline
        $BW_{i}$ & Midhaul bandwidth of $i_{th}$ DU\\
         \hline
        $BW_{max}$ & Maximum midhaul bandwidth available\\
         \hline
        \multirow{2}{*}{$x_i$} & A binary variable which indicates if DU $i$ uses \\
        & Option-7 (\textbf{0}) or Option-6 (\textbf{1})\\
         \hline
        \multirow{2}{*}{W} & A set of bandwidth cut off for each of the \\
        &available split options\\
         \hline
        $w_i$ & Index of the elements in list $W$\\
     \hline   
    \end{tabular}
    \caption{Variables used in the Optimization Model}
    \label{variables}
    \vspace{-0.64cm}
\end{table}

\begin{equation}\label{eq:bandwidths}
W(i) = 
\begin{cases}
133 \text{ Mbps}, & \text{if }\text{$i=1$} \\
9.408\times \mathcal{B}(i-1) \text{ Mbps}, & \text{if }\text{$i>1$}
\end{cases} 
\end{equation} 

The midhaul bandwidth required for each DU $i$, $W(i)$ for a given split option $x_i$ can be calculated by Eqn.(\ref{eq:midhaul}). 

\begin{equation} \label{eq:midhaul}
\quad BW_i = x_i\times UT_i + W(w_i)
\end{equation} 
where
\begin{equation*}
    \scriptstyle \displaystyle{ w_i = 
      x_i + k(1-x_i)\\}
\end{equation*}
and
\begin{equation*}
    \scriptstyle \displaystyle{ k = 1 + 
     \text{index in } \mathcal{B} \text{ to support } UT_i\\}
\end{equation*}

The objective function of our optimization problem that finds the optimal selection of split option for each DU is given below in Eqn.(\ref{eq:obj}) and constraint in Eqn.(\ref{eq:constraint}). The value of $W(w_i)$ and $BW_i$ are given in Eqn.(\ref{eq:bandwidths}) and Eqn.(\ref{eq:midhaul}), respectively. 

%\underline{Objective Function}:
\begin{align}\label{eq:obj}
\displaystyle \max_{x_i,w_i}: \left (\scriptstyle \displaystyle \underbrace{\sum_{i=1}^{n_{DU}}x_i\times UT_i }_{\text{(A)}} + \underbrace{W(w_i)\times(1 + UT_i)}_{\text{(B)}}\right) 
\end{align} 
%\underline{Constraint:}

\begin{equation}\label{eq:constraint}
\scriptstyle \displaystyle \sum_{i=1}^{n_{DU}}BW_i \leq BW_{max}; 
\end{equation}

%\begin{equation}\label{eq2}
%    \scriptstyle \displaystyle{ x_i \in \{0,1\}}
%\end{equation}
%\begin{equation}
%    \scriptstyle \displaystyle{ W = \{W_1, W_2, ..., W_n\}}
%\end{equation}
% Where $W(1) = 133$ and rest of the elements (for $i\geqslant 2$) can be calculated as

% 9.408\times \mathcal{B}(i-1) \text{ Mbps}

% $m_i$ is the index of the elements in $W$ starting from 1.

%There are two main components of the objective function in the proposed optimization model denoted as A and B in Eqn.(\ref{obj}).

\begin{enumerate}
    \item Term A in the objective function maximizes the midhaul bandwidth utilization by considering maximum centralization for the given set of BSs.
    \item Term B in the objective function ensures that the DUs with higher number of  users are considered over DUs with less number of users in order to achieve more centralization. 
\end{enumerate}
The constraint given in Eqn.(\ref{eq:constraint}) ensures that the sum of midhaul bandwidths utilized for each CU-DU pair does not exceed the available midhaul bandwidth in the RAN network.
\begin{figure}[ht]
\centering
\includegraphics[width=0.7\columnwidth]{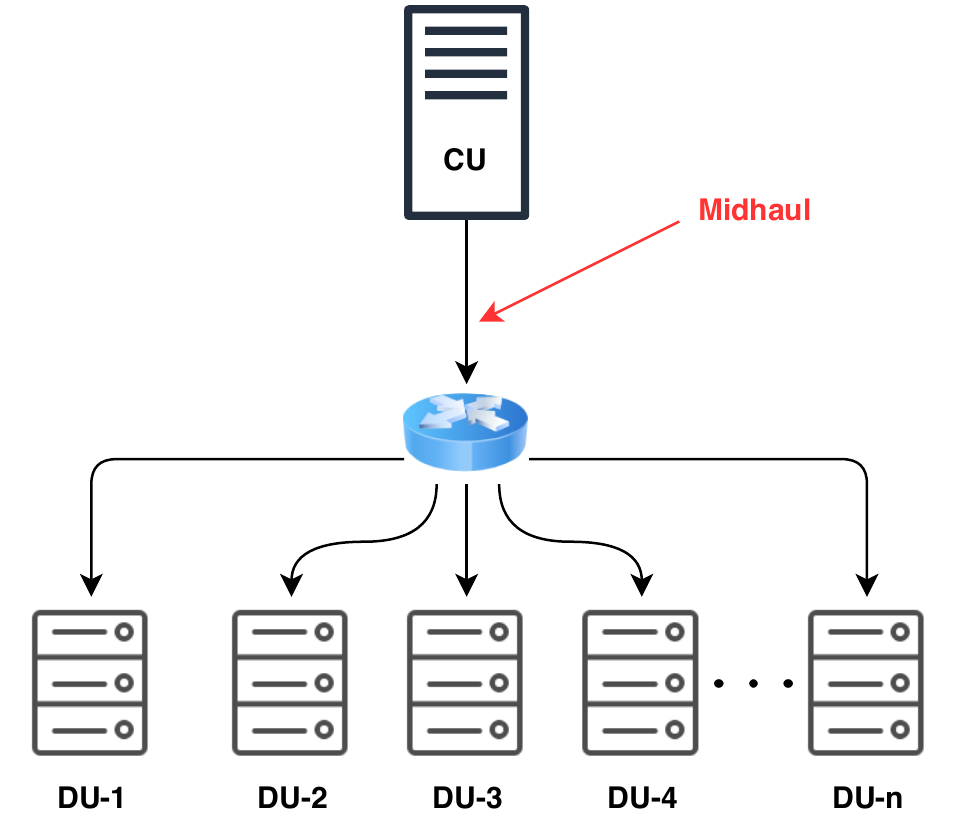}
 \caption {Setting up CU-DU using a shared midhaul network.}
\label{system}
\vspace{-0.7cm}
\end{figure}
%Eqn.(\ref{eq2}) indicates if the DU i uses Option-7 ($x_i = 0$) or Option-6 ($x_i =1$. 
%Midhaul bandwidth requirement for Option-7 does not depend upon user traffic, so we consider Option-7 for five different channel bandwidths in our scenario (i.e., $\mathcal{B}$ = [100,80,60,40,20]). An operator can select any set of bandwidths for a given set of DUs.
This optimization model is flexible enough to consider any set of channel bandwidths that the BS can operate. Apart from deciding the split option for the DUs, the optimization also provides the channel bandwidth to be used when Option-6 is considered for the DUs. 

%In summary, Optimization model dynamically choose the functional splits among six available options, i.e., Option-7 on 20 Mhz,  40 Mhz, 60 Mhz, 80 Mhz, 100 Mhz and Option-6 with 100 Mhz. 
%Eqn.(\ref{eq3}) listings the constant bandwidths for the above six different options available.
%Eqn.(\ref{eq4}) indicates the index of $W$ list.
%Eqn.(\ref{eq5}) gives weight to the DU based on user traffic, and it ensures that a CU-DU pair which is having more number of users runs on a higher split to give the benefits of centralization to more number of users.

\section{Performance Evaluation}
\vspace{-0.15cm}
\begin{figure}[b]
\vspace{-0.62cm}
\centering
\includegraphics[width=\linewidth]{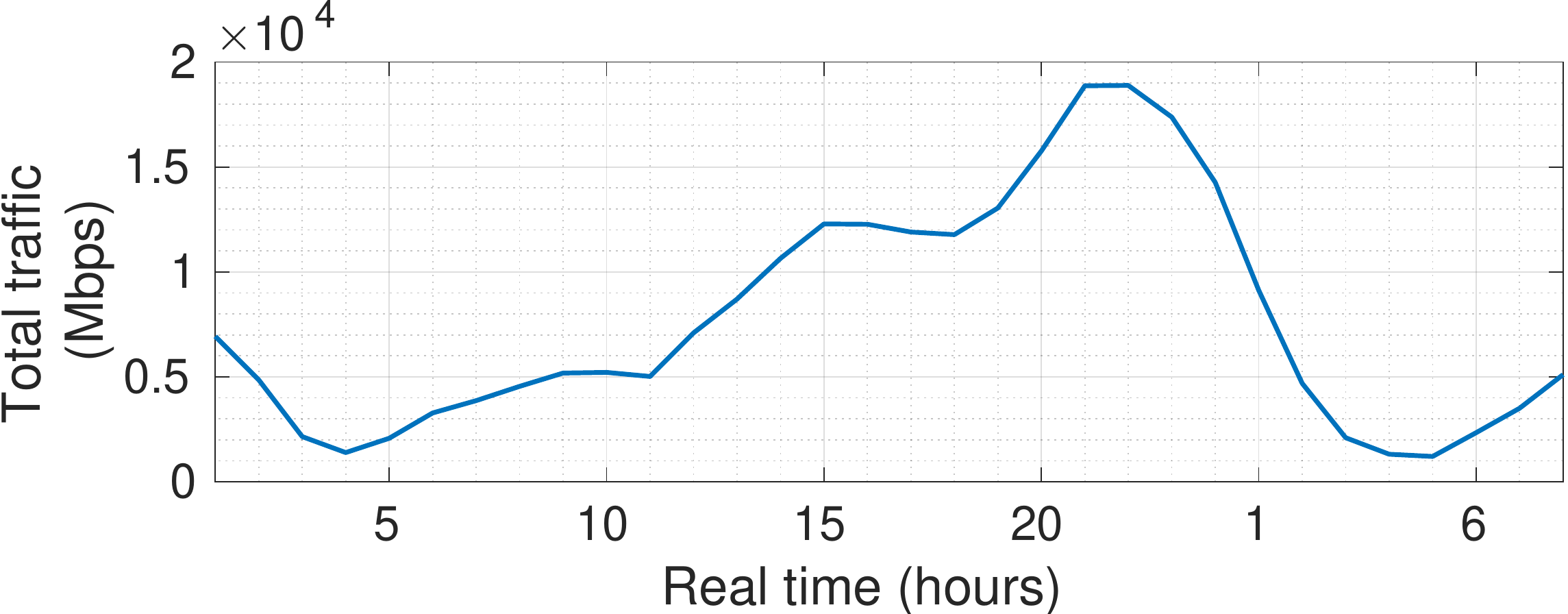}\\
\includegraphics[width=0.99\columnwidth]{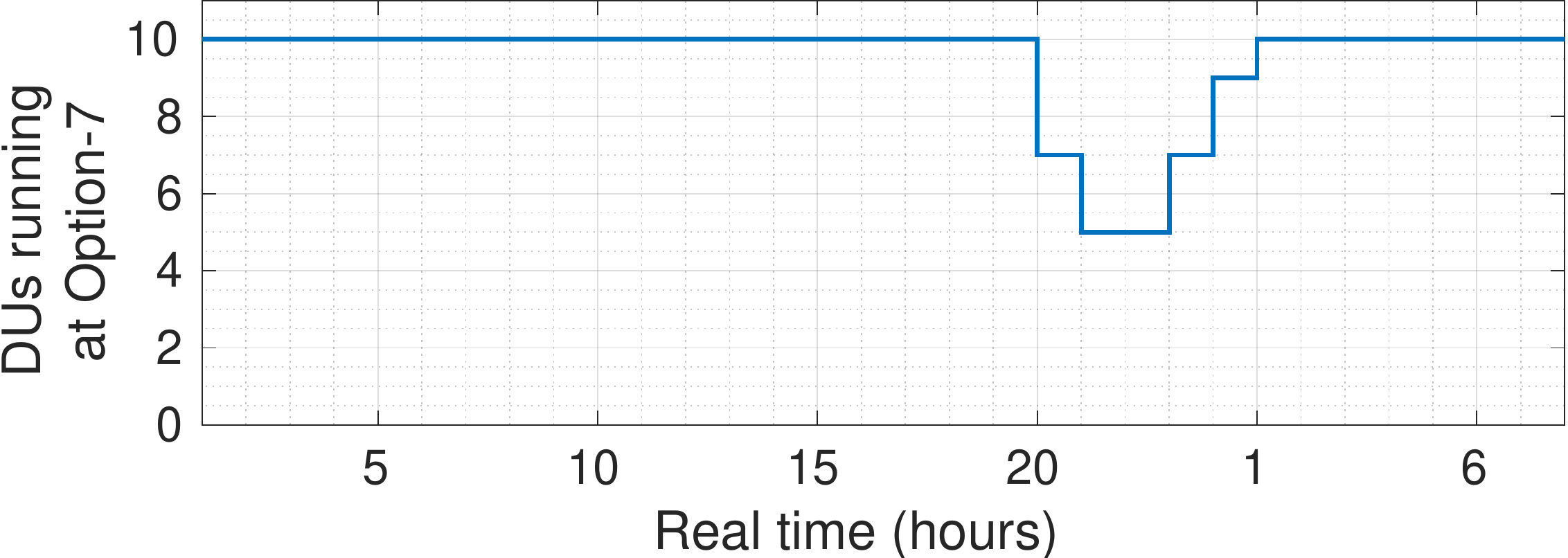}
 \caption {Number of DUs centralized (lower part of figure) w.r.t. variation in total traffic load of DUs (upper part of figure).}
\vspace{-0.4cm}
\label{result1}
\end{figure}
\subsection{Experimental Setup}

\begin{figure*}[ht]

\minipage{0.30\textwidth}
 \centering
\includegraphics[width=\linewidth]{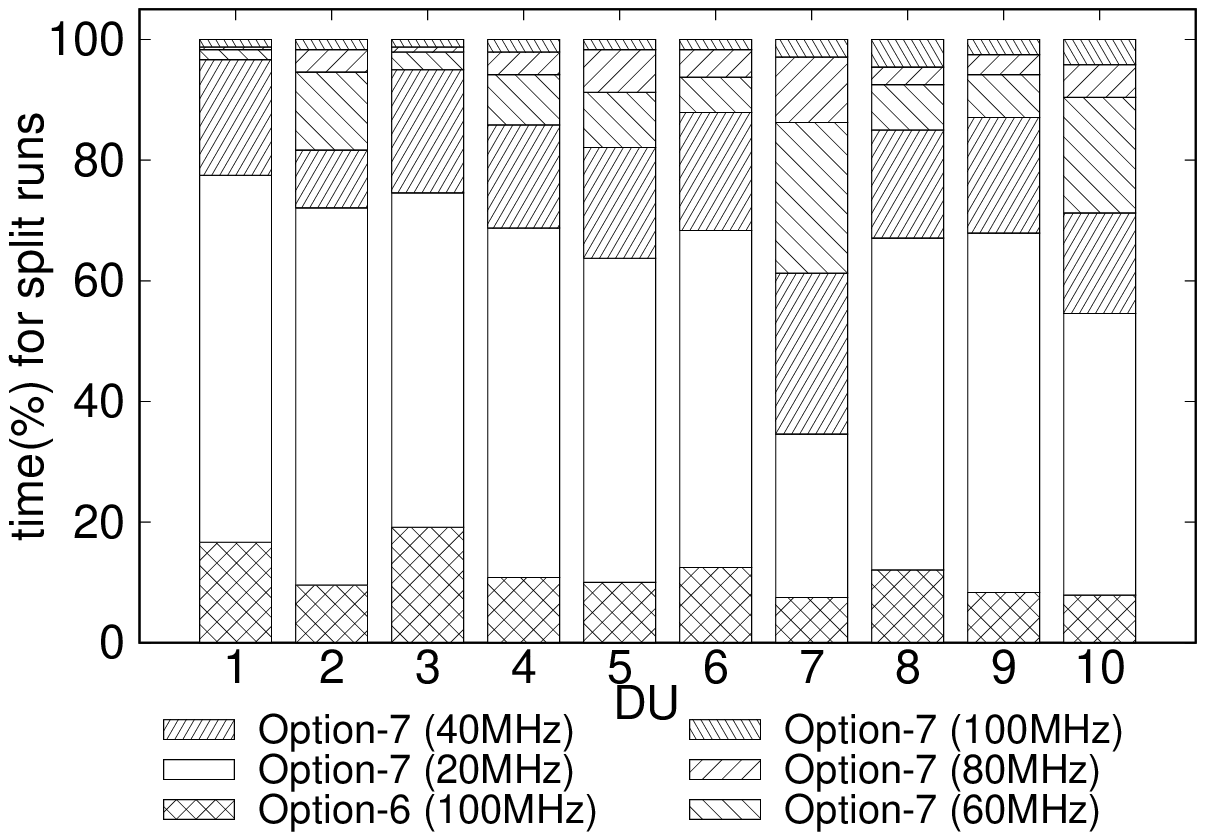}
 \caption {Percentage of the time each DU operates on different functional split options.}
 \label{result2}
\endminipage\hfill
\minipage{0.30\textwidth}
 \includegraphics[width=\linewidth]{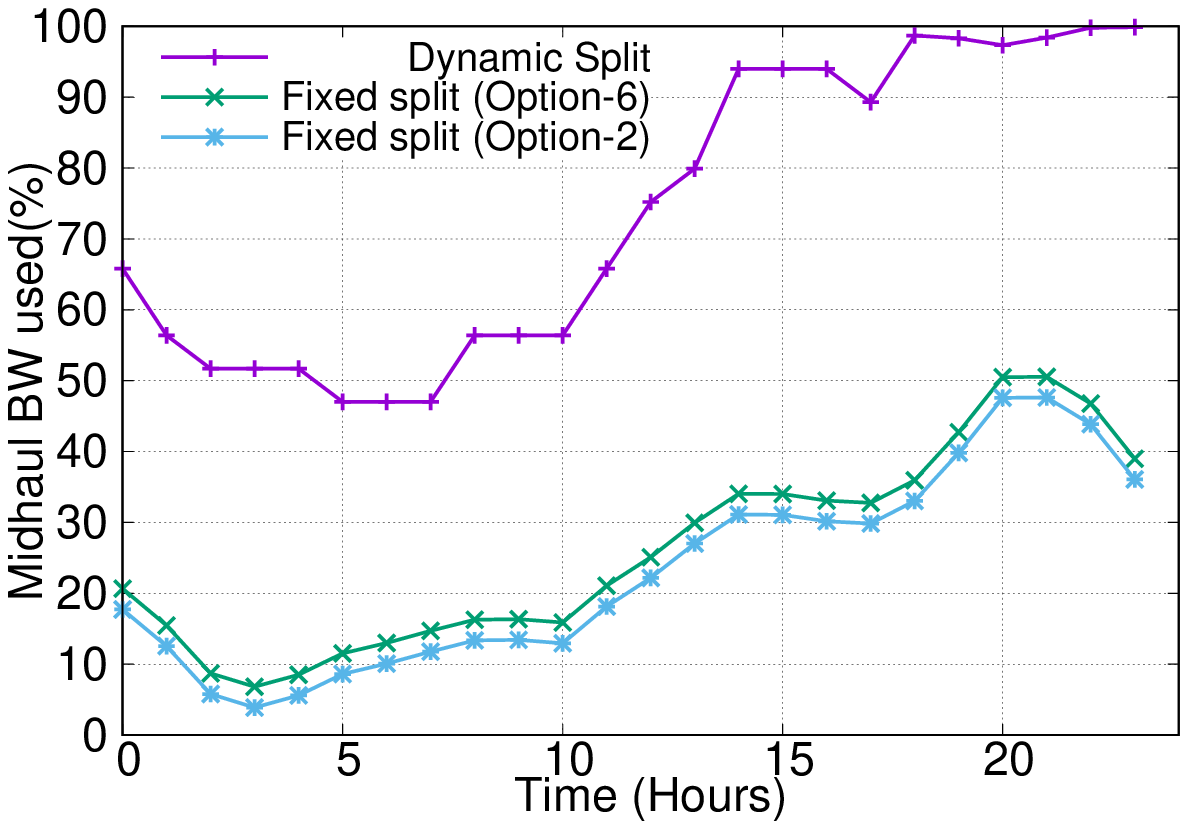}
 \caption {Percentage of midhaul bandwidth used by different functional split options.}
 \label{result3}
\endminipage\hfill
\minipage{0.30\textwidth}%
  \includegraphics[width=\linewidth]{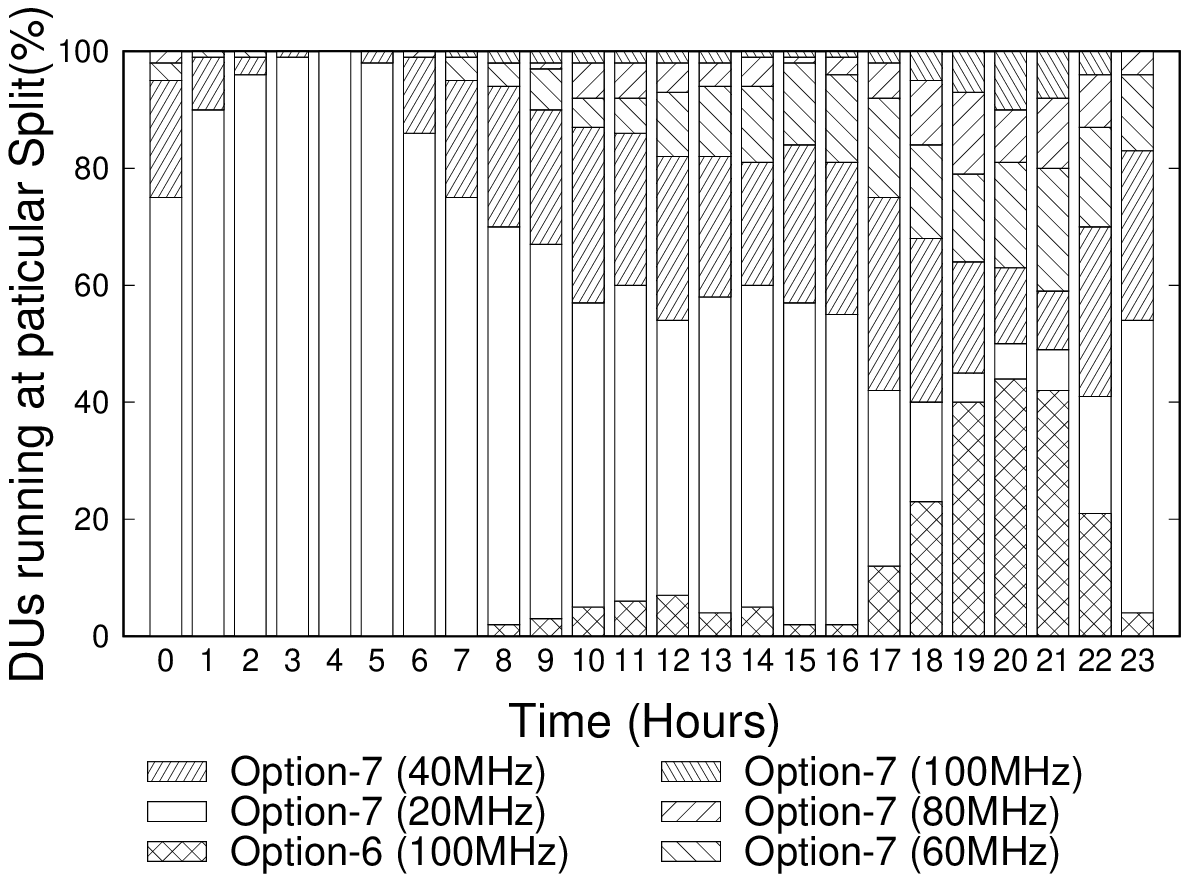}
 \caption {Percentage of DUs operate on different functional split at different time of the day.}
% \vspace{-0.1cm}
\label{result4}
\endminipage
\vspace{-0.64cm}
\end{figure*}

To evaluate the benefits of the proposed dynamic functional split mechanism, we consider a RAN  architecture based on the reference architecture of the 3GPP~\cite{3gpp_38804}. Fig.~\ref{system} shows the considered RAN topology which consists of a CU that connects with the DUs through an optical network connected switch. In our model, we assume that the optical network has a fixed allocated bandwidth between the CU and the DUs which is very much essential for the C-RAN requiring guaranteed bandwidth and delay budget. Moreover, the CU is assumed to possess sufficient computing resource for the split option selected for all the DUs given by the algorithm (\emph{i.e.,} CU computing resource is not a bottleneck here). For a realistic traffic pattern for the load on the DUs, we consider the data collected from a real network available at~\cite{dataset}. The data is collected in an LTE network for approximately $1$ year in $57$ cells at every $1$ hour interval. As the data is collected in an LTE network, we have scaled the data using Eqn.(\ref{scale}) to make it suitable for 5G network. As given in Table~\ref{midhaul}, the maximum data rate of a 5G BS having a $100\ MHz$ channel is considered to be $4\ Gbps$. 

\begin{equation}\label{scale}
     DU_i = \frac{4Gbps}{max\{\text{UserTraffic}_{DU_i} \}} \text{  } \forall \text{   DU} \in \text{Dataset}
\end{equation}

 In the scaled dataset, we have chosen $10$ DUs among the $57$ DUs, that are placed at a different geographical locations following distinct traffic patterns. A subset of DUs are placed near the commercial area while remaining DUs are located in residential colonies. We used the data for these $10$ DUs for $10$ consecutive days to observe the effect of both weekday and weekend traffic patterns. We considered only $5$ channel bandwidths given in Table~\ref{midhaul} and a RAN network similar to Fig.~\ref{system} where all the $10$ DUs are connected to a single CU sharing the midhaul bandwidth. A midhaul bandwidth of $41\ Gbps$ is considered which is sufficient to run all the $10$ DUs in Option-6 with $100\ MHz$. Hence, we do not scrutinize Option-2 with $100\ MHz$ in our experiment. The main benefit of the dynamic function split is to use the tidal traffic pattern of the BSs and optimally use the available midhaul bandwidth. For the simulation, we solve the optimization problem for every time epoch of traffic data, to find the optimal functional split for each of the $10$ DUs. We present the results obtained over $30$ simulation \cite{git} trials.

% \subsection{Dataset}
% \begin{itemize}
%     \item The data is collected for approximately 1 year x 24 hours x 57 cells.
%     \item Sample interval is 1-hour.
%     \item Data is available for LTE, so we scaled it to use in our 5G scenario.
%     \item Scaling factor for $ DU_i = \frac{4Gbps}{max\{UserTraffic_{DU_i} \}}$.
%     \item We picked 10 different DUs placed at different geographical locations following different traffic patterns for simulations.
%     \item The data for these DUs is chosen for 10 consecutive days.
% \end{itemize}

%\section{Results}

% and have collected the results with a confidence interval of $99\%$, but the value of the error bars are barely visible in the plot.

\vspace{-0.19cm}
\subsection{Centralization benefits of dynamic functional split}
\vspace{-0.15cm}
Fig.~\ref{result1} shows the number of DUs centralized with varying traffic load (sum of downlink traffic of $10$ DUs) for a period of $32$ hours. We observe that, all the DUs are able to achieve maximum centralization (using Option-7) in low and moderate traffic conditions. During high traffic load, $6$ DUs are running on Option-7 with different channel bandwidth while other DUs are running Option-6 with $100\ MHz$ channel bandwidth.If the operator chooses Option-7 $100\ MHz$  statically for 10 DUs, it costs operator $90\ Gbps$ midhaul bandwidth, which is only needed during the high traffic load. Consequently, the operator underutilizes the network resources during low and moderate traffic loads.

\vspace{-0.19cm}
\subsection{Percentage  of  time  each  DU  operates  on  different functional split options}
\vspace{-0.15cm}
Fig.~\ref{result2} shows the percentage of time each DU operates on different split options over a $10$ days period for given midhaul bandwidth. Note that, about $60\%$ of the time, each DU is able to operate on Option-7 with $20\ MHz$ channel bandwidth, \emph{i.e.,} each DU is achieving the maximum centralization without compromising on meeting the required BS capacity for a given midhaul bandwidth. We could not achieve this maximum centralization for $60\%$ of the time if we have considered the Option-7 with a fixed channel bandwidth of $100\ MHz$. 
\vspace{-0.19cm}
\subsection{Percentage  of  midhaul  bandwidth  used  by  different functional split options.}
\vspace{-0.15cm}
Fig.~\ref{result3} compares and contrasts among dynamic functional split, fixed functional split (Option-6), and fixed functional split (Option-2) with respect to percentage of midhaul bandwidth used at different time of a given day. We observe that, dynamic split uses available bandwidth optimally while Option-6 and Option-2 use maximum $50\%$ of the midhaul bandwidth only during the peak traffic hours. We haven't shown the fixed functional split (Option-7), because the available midhaul bandwidth is not sufficient to handle all the DUs.  
\vspace{-0.5cm}
\subsection{Percentage of DUs operate on different functional split options at different time of the day}
\vspace{-0.15cm}
Fig.~\ref{result4} depicts the percentage of DUs among the $10$ DUs in our experiment operating on each of the functional split options at every hour of the day. We can observe that, during the early hours (12 AM - 9 AM), most of the DUs operate on Option-7 $20\ MHz$ channel bandwidth as the traffic loads on the DUs are very low during this period. During medium traffic load (10 AM - 4 PM), most of the DUs operate on both Option-7 $40\ MHz$ and Option-7 $20\ MHz$ to meet the traffic requirement at each of the BSs. During the peak load (6 PM - 10 PM), only half of the DUs operate on Option-7 $40\ MHz$ and Option-7 $60\ MHz$ variations while remaining DUs operate on Option-6 to meet the traffic demand and with the limited midhaul bandwidth.
% \begin{figure}[h]
% \centering
% \includegraphics[width=0.9\columnwidth]{6_Hourly_analysis_of_Dynamic_Split.eps}
%  \caption {Percentage of DUs operate on different functional split options at different time of the day.}
% % \vspace{-0.1cm}
% \label{result4}
% \end{figure}
\vspace{-0.19cm}
\subsection{Percentage of time each functional split options were used for a 10 days period}
\begin{figure}[t]
\centering
\includegraphics[width=0.9\columnwidth]{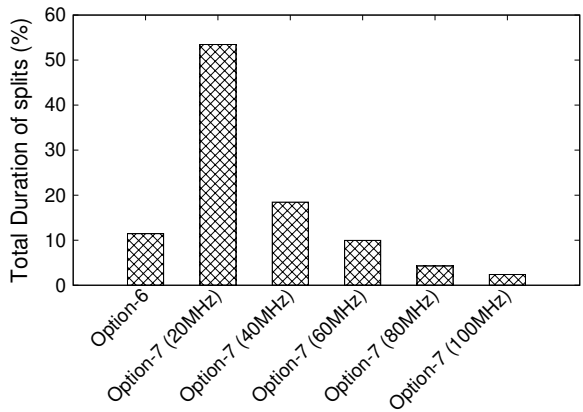}
 \caption {Percentage of time each functional split options were used over 10 days period.}

\label{result5}
\vspace{-0.65cm}
\end{figure}

\vspace{-0.15cm}
Fig.~\ref{result5} shows average percentage of the time different functional split options are used by $10$ DUs over $10$ days period. Option-7 with $20\ MHz$ channel bandwidth is used for more than $50\%$ of the time as we can observe from the traffic pattern in the dataset that peak traffic load for the BSs occur only for a small duration in a given day. We can see that, $90\%$ of the time, the proposed dynamic functional split mechanism is able to  maximize the centralization by operating them on Option-7 with  one of the available channel bandwidths $20\ MHz$, $40\ MHz$, $60\ MHz$, and $100\ MHz$.
\vspace{-0.2cm}
\section{Related Work}
\vspace{-0.15cm}
In the literature, trade-off between different functional splits are explored thoroughly. In~\cite{Himank}, the author did the experimental evaluation of the impact of virtualizing eNB functions on the fronthaul latency and jitter budget when different virtualisation methods are utilized. Their results showed that lighter virtualization methods (e.g., Docker) are impacting the fronthaul latency budget for Option 7-1 (\emph{i.e.,} intra-PHY) split less than heavier virtualization methods (e.g., VirtualBox). However, in all the cases, the fronthaul latency budget reduction depends on the considered signal bandwidth. The higher the bandwidth the higher the computations required and the higher the fronthaul latency budget reduction. The authors of \cite{survey} provide a comprehensive literature overview of the functional split options proposed by the 3GPP. Each functional split has been discussed in a detailed description of the location and abilities. This gives insight on what is being transmitted on the fronthaul link but also which functions are located in the DU and the CU, respectively. This also gives the impact of the chosen functional split on the fronthaul network connecting the DU to the CU-pool both in terms of fronthaul bitrates and latency.
The authors in~\cite{8408566}, proposed Flex5G which selects the appropriate functional split for different small cells utilizing midhaul bandwidth and minimizing the inter cell interference. The authors in~\cite{8968382} proposed a flexible function split between CU and DUs for delay critical applications and showed the impact of delay constraints on the required midhaul bandwidth and the power consumption. Their simulation results demonstrate that delay constraint has a significant impact on the required fronthaul bandwidth and power consumption. 

In~\cite{Himank}, the authors studied the effect on the latency budget available for the midhaul caused by the choice of functional split option and virtualization technology used in a virtualized C-RAN. All these works assume that either functional split between the CU and DU is fixed at the time of network deployment because the functional split for a given CU DU pair is based on the available midhaul bandwidth and the delay budget requirement of the specific functional split used. The authors of \cite{8845147} presented an adaptive 5G RAN implementation that supports migrations between functional splits at run-time. They described the challenges related to service disruption due to split migration at run time. Then a proposal to switch from MAC-PHY to PDCP-RLC without service interruption is given.
\vspace{-0.25cm}
\section{Conclusions}
\vspace{-0.15cm}
In this paper, we proposed a dynamic functional split mechanism for 5G C-RAN to effectively use the midhaul bandwidth and achieve maximum centralization gains by tuning the channel bandwidth optimally. Optimal functional split selection for a given traffic load for the DUs is obtained using the proposed optimization model. We achieve 90\% of centralization using our proposed mechanism, encompassing the channel bandwidth. This mechanism opens up a lot of other problems to be solved in the future such as how to switch between functional splits for a given cell site in real-time and cost-benefit analysis of different split options.
\vspace{-0.250cm}
% % \section*{Acknowledgment}
% \section{TO UPDATE}

% The variables used in the objective function are listed below in Table ~\ref{variables}.
% \\
% \underline{Objective Function}:
% \begin{align}\label{obj}
% \displaystyle \max_{x_i,w_i}: \left (\scriptstyle \displaystyle \underbrace{\sum_{i=1}^{n_{DU}}x_i\times UT_i }_{\text{(A)}} + \underbrace{W(w_i)\times(1 + UT_i)}_{\text{(B)}}\right) 
% \end{align} 
% \underline{Constraints :}
% \begin{equation}\label{eq1}
% \scriptstyle \displaystyle \sum_{i=1}^{n_{DU}}BW_i \leq BW_{max}; \quad BW_i = x_i\times UT_i + W(w_i)
% \end{equation}
% \begin{equation}\label{eq2}
%     \scriptstyle \displaystyle{ x \in \{0,1\}}
% \end{equation}
% \begin{equation}\label{eq3}
%     \scriptstyle \displaystyle{ W = \{133, 9400, 7520, 5640, 3760, 1880\}}
% \end{equation}
% \begin{equation}\label{eq4}
%     \scriptstyle \displaystyle{ w_i \in [1,2,3,4,5,6]}
% \end{equation}
% \begin{equation}\label{eq5}
%     \scriptstyle \displaystyle{ w_i(UT_i)= \begin{cases} 
%       x + 2(1-x) & 3.2 \leq UT_i \\
%       x + 3(1-x) & 2.4 \leq UT_i < 3.2 \\
%       x + 4(1-x) & 1.6 \leq UT_i < 2.4 \\
%       x + 5(1-x) & 0.800 \leq UT_i < 1.6 \\
%       x + 6(1-x) & UT < 0.800 \\
%   \end{cases}}
% \end{equation}

\ifCLASSOPTIONcaptionsoff
  \newpage
\fi
\bibliographystyle{IEEEtran}
\vspace{-0.03cm}
\bibliography{biblio.bib}

\end{document}